\documentclass[final]{svjour3}
\usepackage{graphicx}
\usepackage{verbatim}
\usepackage{rotating}
\usepackage{amssymb}
\usepackage{mathptmx}
\usepackage[sort&compress,numbers]{natbib}
\usepackage{doi}%<----------
\usepackage{hyperref}
\makeatletter
\journalname{Journal of Low Temperature Physics}
\bibpunct{}{}{,}{s}{}{,}

\begin{document}

\newcommand{\hdblarrow}{H\makebox[0.9ex][l]{$\downdownarrows$}-}
\title{The BeEST Experiment: Searching for Beyond Standard Model Neutrinos using $^7$Be Decay in STJs}

\author{K.G.~Leach$^\dagger$~~and~~S.~Friedrich$^\ddagger$\\for the BeEST Collaboration}

\institute{$\dagger$Department of Physics, Colorado School of Mines\\ Golden, CO, 80401 USA\\
\email{kleach@mines.edu}\\\\
$\ddagger$Nuclear and Chemical Sciences Division, Lawrence Livermore National Laboratory\\ Livermore, CA 94550, USA\\
\email{friedrich1@llnl.gov}}

\maketitle

\begin{abstract}
Precision measurements of nuclear $\beta$ decay are among the most sensitive methods to probe beyond Standard Model (BSM) physics in the neutrino sector.  In particular, momentum conservation between the emitted decay products in the final state is sensitive to any massive new physics that couples to the neutrino mass.  One way to observe these momentum recoil effects experimentally is through high-precision measurements of electron-capture (EC) nuclear decay, where the final state only contains the neutrino and a recoiling atom.  The BeEST experiment precisely measures the eV-scale radiation that follows the radioactive decay of $^7$Be implanted into sensitive superconducting tunnel junction (STJ) quantum sensors.  STJs are ideally suited for measurements of this type due to their high resolution at the low recoil energies in EC decay, and their high-rate counting capabilities.  We present the motivation for the BeEST experiment and describe the various phases of the project.

\end{abstract}

\section{Introduction}

The lepton sector of the Standard Model (SM) provides perhaps the best window into Beyond Standard Model (BSM) physics given the confirmed observation of non-zero neutrino masses~\cite{Ahm01,Fuk98}, the recent confirmation of the muon $g-2$ anomaly~\cite{Abi21}, and several other outstanding claims of leptonic BSM physics~\cite{Gar20}.  As a result, extensions to the SM description of leptons are unavoidable and must at least account for the fact that neutrinos have non-zero mass eigenstates.  So-called ``sterile neutrinos" are are present in almost all SM extensions that include neutrino mass, and have the characteristic property that they are non-interacting with respect to the SM forces.  These sterile neutrinos are associated with heavy mass states that do not depend on the Higgs mechanism and can therefore be on nearly any mass scale.  keV-scale sterile neutrinos are among the most highly motivated, since they have the right cosmological properties to explain the observed dark matter in our universe~\cite{Boy19,Adh17}.  In fact, extensions that include neutrinos in this mass range, such as the $\nu$MSM, are able to provide a framework that can simultaneously explain the observational data for dark matter and baryon asymmetry of the Universe~\cite{Asa05a,Asa05b}.

\begin{figure}[t!]
\centering
\includegraphics[width=\linewidth]{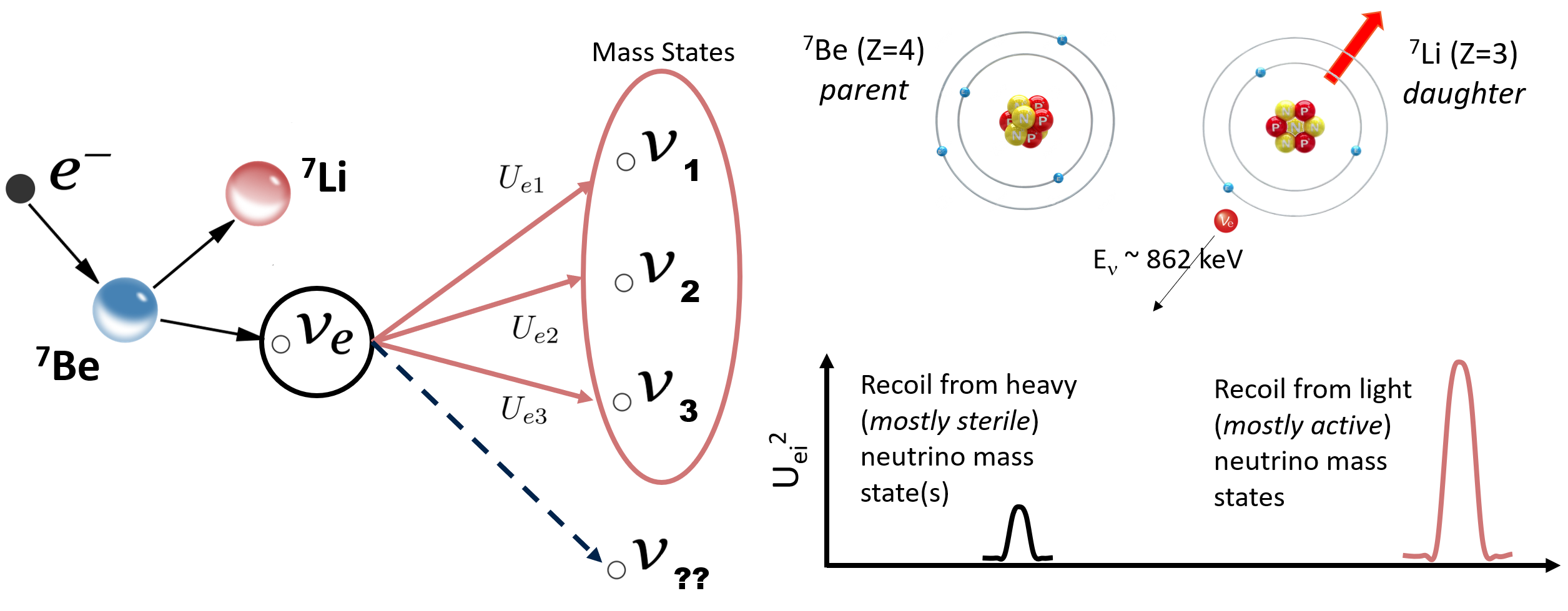}
\caption{\label{processes}A schematic of the nuclear EC decay of beryllium-7 (left) illustrating the superposition of neutrino mass states ($\nu_{1,2,3}$) that exist for the electron neutrino ($\nu_e$) -- and the potential for heavy BSM mass states ($\nu_{??}$).  The measured lithium-7 recoil energies measured in the experiment differ depending on the mass-state of the emitted neutrino, and is schematically shown in the bottom right.}
\end{figure}

Effective experimental searches for these particles should be model-independent and cover a large area of the allowed parameter space.  The most powerful approach to search for these heavy neutral leptons (HNLs) in the sub-MeV mass range is through energy and momentum conservation in weak nuclear decay.  The pure electron capture (EC) decaying nucleus $^7$Be is the ideal case for neutrino studies using this approach due to its pure two-body final state, its relatively large decay energy, and simple atomic and nuclear structure.  Since the quantum information of the $\nu_e$ and recoiling $^7$Li daughter atom are entangled following EC decay, the superposition of mass eigenstates that make up the $\nu_e$ flavor state can be probed through precision measurements of the atomic recoil momentum.  All possible neutrino mass eigenstates (including any outside of $\nu_{1,2,3}$) continue to exist as a superposition until the wavefunction is collapsed through detection of the $^7$Li.  When the wavefunction collapses, a single mass state is selected and the observed kinetic energy of the recoiling atom differs depending on the mass of any heavy particle that is emitted.  The fraction of events that are shifted with respect to the SM signal determines the new physics mixing fraction with $\nu_e$ (Fig.~\ref{processes}).  This method of decay momentum reconstruction is a simple, model-independent approach to HNL searches since it relies only on the existence of a heavy neutrino admixture to the active neutrinos and not on the model-dependent details of their interactions.

\section{\label{ecdecay}Nuclear Electron Capture Decay of $^7$Be}
$^7$Be decays by EC primarily to the nuclear ground-state of $^7$Li with a $Q$ value of 861.89(7)~keV~\cite{AME16} and a half-life of $T_{1/2}=53.22(6)$~days~\cite{Til02}.  A small branch of 10.44(4)\% results in the population of a short-lived excited nuclear state in $^7$Li ($T_{1/2}=72.8(20)$~fs)~\cite{Til02} that de-excites via emission of a 477.603(2)~keV $\gamma$-ray~\cite{Hel00}.  In the EC process, the electron can be captured either from the $1s$ shell ($K$-capture) or the $2s$ shell of Be ($L$-capture). For $K$-capture, the binding energy of the $1s$ hole is subsequently liberated by emission of an Auger electron whose energy adds to the decay signal and separates it from the $L$-capture signal.  Since the nuclear decay and subsequent atomic relaxation occur on short time scales, a direct measurement produces a spectrum with four peaks: two for $K$-capture and two for $L$-capture into the ground state and the excited state of $^7$Li, respectively.  Due to the relative spatial overlaps of the $1s$ and $2s$ electron orbitals with the nucleus, the EC process is dominated by $K$-shell capture with a measured $L/K$ capture ratio of 0.040(6) in HgTe~\cite{Voy02} and 0.070(7) in Ta~\cite{Fre20}.   The simplicity of the atomic structure in particular is one of the primary reasons for using $^7$Be to perform these BSM physics searches.

\section{Superconducting Tunnel Junctions (STJs)}
\begin{figure}[t!]
\centering
\includegraphics[width=\linewidth]{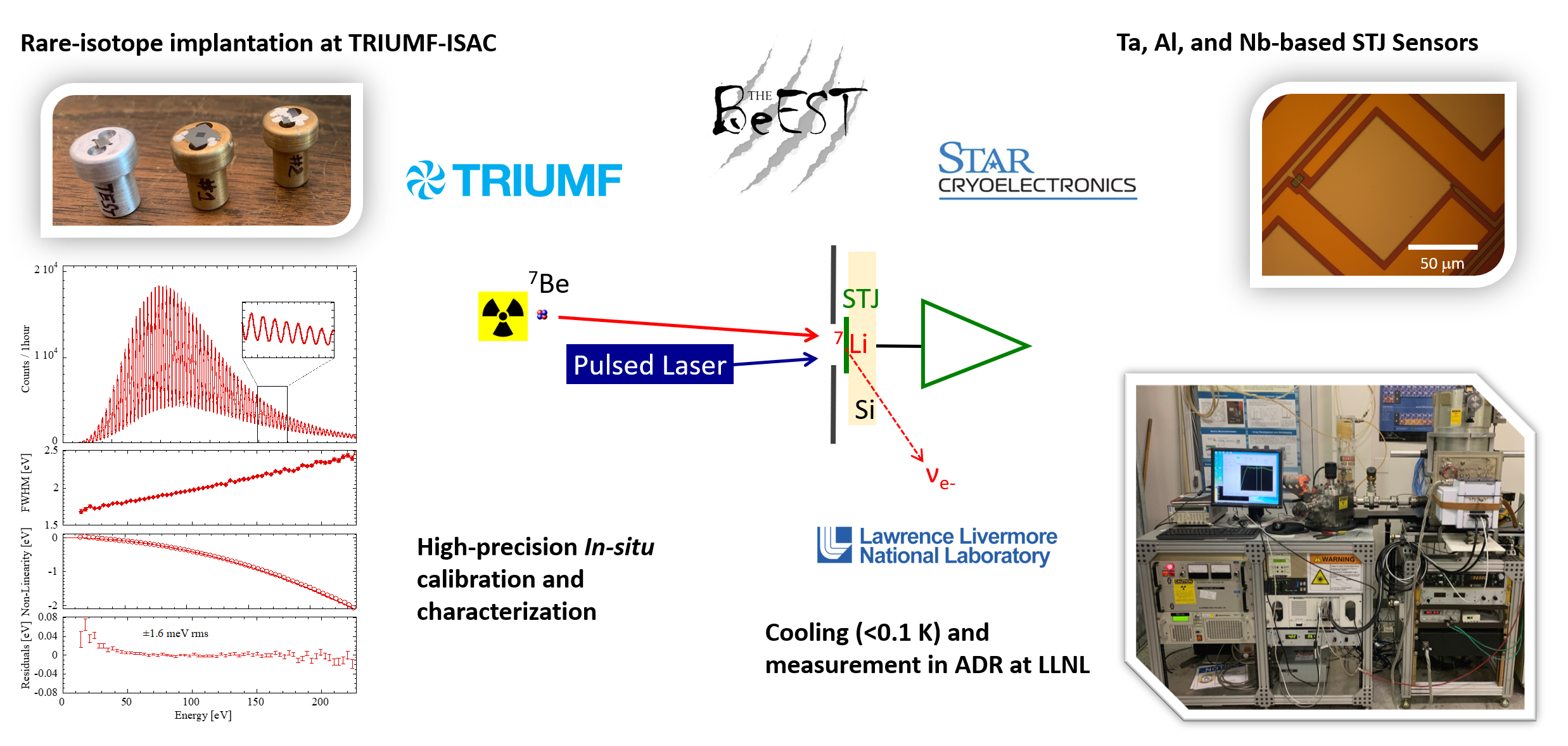}
\caption{\label{BeEST}A schematic of the BeEST experimental concept.}
\end{figure}
The detection and characterization of $\sim 20 - 120$~eV recoil energies from $^7$Be EC decay at the precision required to provide competitive limits on BSM physics are difficult to achieve.  The BeEST confronts these challenges using high-rate STJ radiation detectors and is the first instance of adapting them to search for BSM physics in nuclear decay.  STJs are high-rate quantum sensors that were originally developed for high-resolution X-ray spectroscopy~\cite{Kur82,Fri03,Ver08} and have been used for many years in astronomy~\cite{Ver08} and at synchrotron facilities~\cite{Car14}.  STJs are a type of Josephson junction that consists of two superconducting electrodes separated by a thin insulating tunneling barrier. The absorption of radiation in one of the electrodes breaks the Cooper pairs of the superconducting ground state and excites free excess charge carriers above the superconducting energy gap $\Delta$ in proportion to the absorbed energy.  As these excess charges tunnel across the barrier, they generate a temporary increase in current that can be directly read out with a field-effect transistor based preamplifier at room temperature. The high energy resolution in STJs is due to the fact that the energy to create an excess charge $\epsilon=1.7\Delta$ scales with the energy gap $\Delta$~\cite{Kur82}. For superconductors, $\Delta$ is of order 1~meV and thus roughly three orders of magnitude smaller than the band-gap in semiconductors, giving rise to $\sim1$~eV resolution at the signal energies relevant to the BeEST~\cite{Pon16,Fri20}.  In order to prevent thermal excitations across these very small gaps, STJs must be operated well below the superconducting transition temperature $T_C$.

The primary advantages of using STJs for this type of experiment are two-fold:
\begin{enumerate}
    \item The maximum STJ count rates are determined by the time that the excess charges remain excited above $\Delta$ before they recombine into the superconducting ground state and again form Cooper pairs, which is typically in the $10~\mu$s range~\cite{Fri09}.  This enables each STJ detector pixel to operate at rates up to $10^4$ counts/s~\cite{Fri03}, which places them among the highest-rate LTD technologies with high energy resolution, and makes them ideal to search for heavy neutrinos with very weak couplings.
    \item The high energy resolution of STJs can be well characterized for signal energies in the range of interest for the BeEST with high-rate pulsed lasers~\cite{Fri20}.  Their response is well understood, very predictable, and can be measured \textit{in-situ}~\cite{Pon16,Fri20,Fri21}.
\end{enumerate}

\section{\label{7Be}Harvesting High-Intensity $^7$Be$^+$ Beams}
For the BeEST experiment, a pure, high-intensity radioactive $^7$Be source is directly implanted into the STJs to ensure that all of the decay products deposit their energy in the devices.  This prevents partial energies from being recorded due to surface effects where the Auger electrons could leave the device with some kinetic energy, thus broadening the observed recoil signals at low energies.  The required $^7$Be is harvested at the TRIUMF Isotope Separator and Accelerator (ISAC) rare-isotope beam facility on the campus of the University of British Columbia in Vancouver, Canada~\cite{Dil14}.  TRIUMF-ISAC is able to produce among the purest ($>99.99\%$), high-intensity ($\geq 10^8$~s$^{-1}$) beam of $^7$Be in the world  via the isotope separation on-line (ISOL) technique~\cite{Blu13}.  To perform the implantation, the low-energy ISAC implantation station (IIS) is used, where laser-purified radioactive ion beams are delivered at energies up to 50~keV for implantation.  The ISAC facility is currently undergoing target development towards increased yields of $^7$Be using a new state-of-the-art graphite ISOL target.  Since the $^7$Be is produced mainly by spallation on carbon, the prospect of using pure graphite foils is attractive since they are exempt from the same license restrictions of running proton currents on uranium-carbide. Thus with a pure graphite target and a proton current of $100~\mu$A, $^7$Be intensities of $\geq10^{10}$~s$^{-1}$ are projected, making the beam-time requirements for Phase-IV of the BeEST less demanding.

\section{The BeEST Experiment}
\begin{figure}[t!]
\centering
\includegraphics[width=0.26\linewidth]{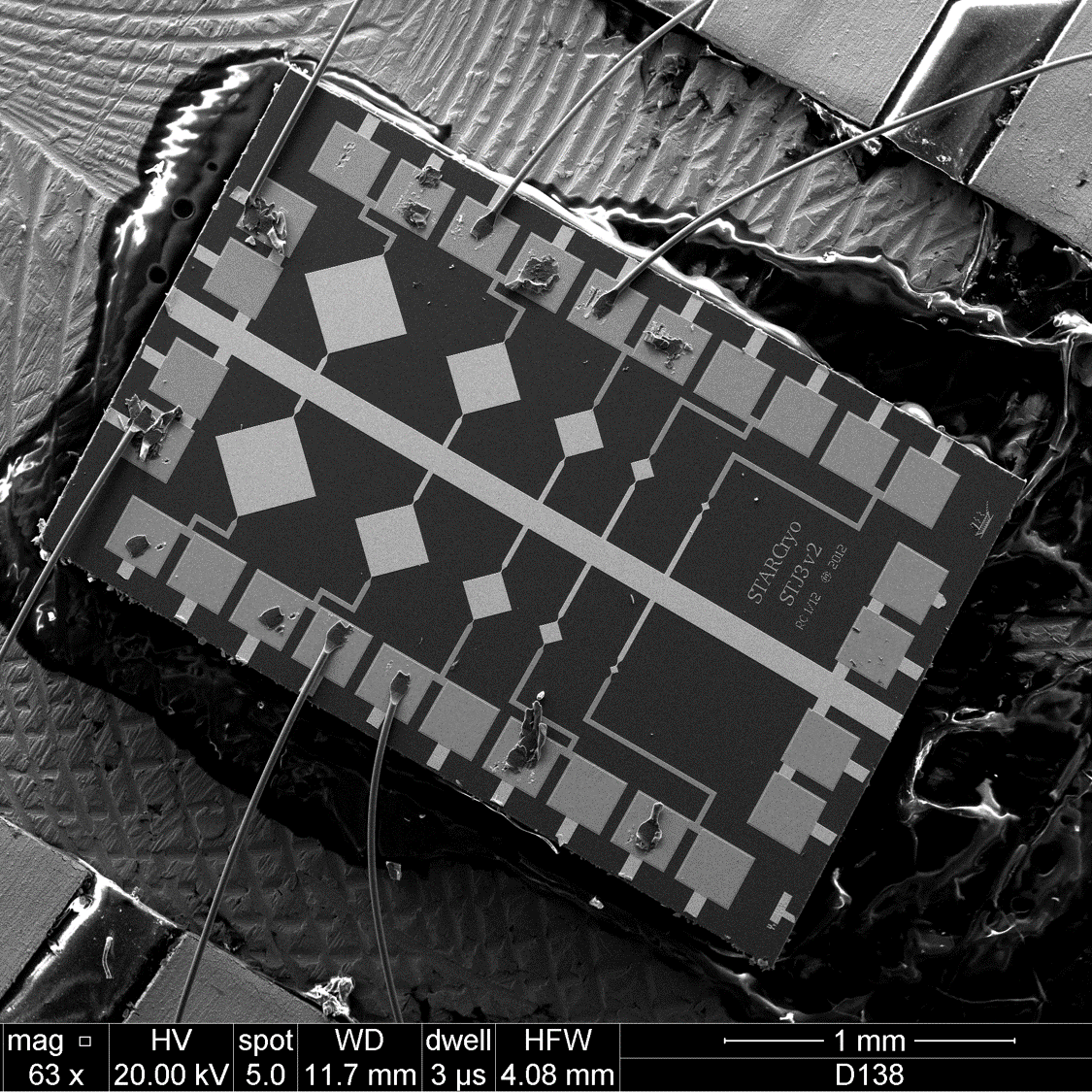}~~~\includegraphics[width=0.345\linewidth]{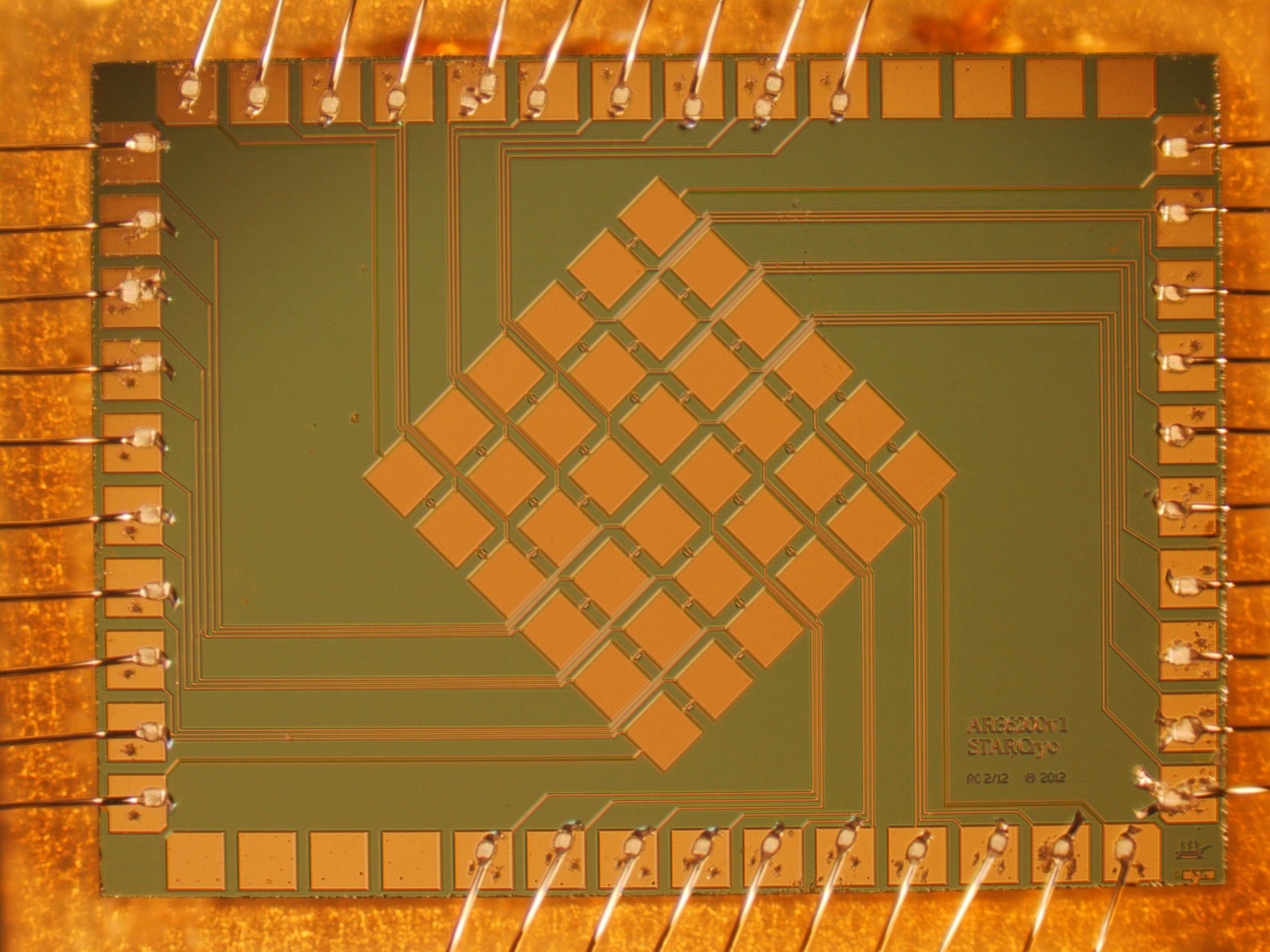}~~~\includegraphics[width=0.27\linewidth]{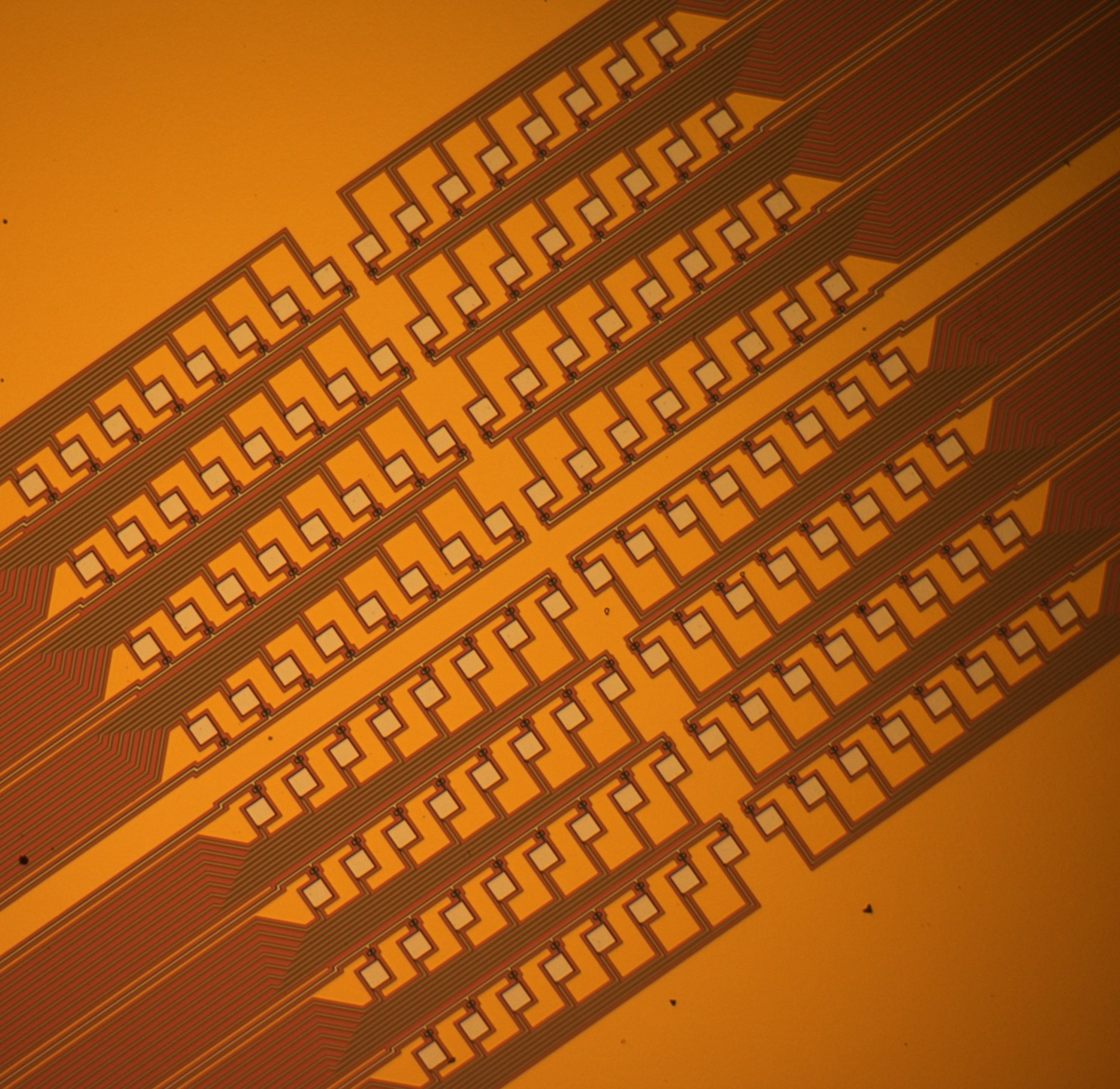}
\caption{\label{Detectors}Images of the detectors for the various phases of the BeEST experiment.  From left to right - a 10-pixel Ta-based chip used for Phases-I and -II, a 36-pixel Ta array used for Phase-III, and a 128-pixel Al-based STJ array prototype for Phase-IV.}
\end{figure}
The BeEST experiment employs a staged approach that employs four phases to demonstrate the concept, evaluate systematics of the experiment, scaling the experiment to multi-pixel arrays, and operation of STJ arrays with new superconducting material.  Phases-I and -II have already been successfully completed~\cite{Fre20,Fri20,Fri21}, and Phases-III and -IV are currently underway.  Briefly described below are the main objectives and milestones for the 4 phases of the BeEST experiment.

\subsection{\label{completed}Phase-I: Proof-of-Concept}
The primary goal of Phase-I was to demonstrate the concept of the BeEST and determine if STJ detectors could be used to perform high-rate nuclear recoil measurements of EC decay.  The Ta-based STJs used were five-layer devices consisting of Ta (165~nm) - Al (50~nm) - Al$_2$O$_3$ (1~nm) - Al (50~nm) - Ta (265~nm) that were fabricated by photolithography at STAR Cryoelectronics~\cite{Car13} (Fig.~\ref{Detectors} (left)).  These detectors were used previously for low-energy decay measurements~\cite{Pon18} and were exceptionally well characterized prior to $^7$Be implantation~\cite{Pon16}. The first $A=7$ irradiation at TRIUMF took place over a 16 hour period where a total of $2\times10^{12}$ $^7$Be$^+$ and $^7$Li$^+$ ions were implanted into 6 STJ detectors at an energy of 26~keV.  Of these ions, approximately $8\times10^{10}$ were implanted into the $(68~\mu$m)$^2$ STJs of interest for Phase-I - requiring each pixel to count at roughly 800 counts/s.  The results of the high-rate Phase-I data demonstrated the concept, and provided the platform to initiate Phase-II of the BeEST.  However, given the large number of $^7$Li atoms that were present in the beam, damage effects were observed in the STJ that required mitigation in Phase-II.

\subsection{\label{PhaseII}Phase-II: First BSM Physics Search and Improved Detector Characterization}
The primary goal of Phase-II was to provide the first exclusion limits in our search for BSM physics in the neutrino sector.  To achieve this, we required the ability to i) evaluate possible ion-beam damage effects suffered in Phase-I, ii) better evaluate the detector response to nuclear recoils, and iii) provide a high-precision energy calibration of the recoil spectrum.  The implantation time was limited to 30 minutes to reduce damage effects, which resulted in $2\times10^{8}$ $^7$Be$^+$ ions implanted into the (138~$\mu$m)$^2$ STJ used for the characterization tests, resulting in an initial event rate per pixel of only $\sim 10$~counts/s. Data were acquired for $\sim$20~hours/day over a period of four weeks in the LLNL ADR.  Since we implanted $100\times$ less beam in the second implantation, the (138~$\mu$m)$^2$ STJs were used, which gave the best combination of statistics and resolution.  For energy calibration of the Phase-II data, the STJs were simultaneously exposed to 3.49965(15)~eV photons from a pulsed Nd:YVO$_4$ laser triggered at a rate of 100~Hz.  We adjusted the laser intensity such that multi-photon absorption provides a comb of peaks over the energy range of interest from 20 - 120 eV in the (138~$\mu$m)$^2$ STJ.  The energy calibration of the spectrum was statistics limited, and resulted in a precision of 0.005~eV - thus providing a successful demonstration of the calibration method for future phases of the experiment. This high-level of detector characterization~\cite{Fri20} allowed for the extraction of exclusion limits from the Phase-II data that were up-to an order of magnitude better than from all previous laboratory-based limits in the 100 - 860 keV mass range~\cite{Fri21}.

\subsection{\label{PhaseIII}Phase-III: Scaling of the BeEST to 36- and 112-Pixel Arrays}
To provide significant increases in the statistical precision of the BeEST experiment, large arrays of STJs are required.  For Phase-III we are currently in the process of evaluating scaling the experiment to 36- and 112-pixel detector arrays of $(200~\mu$m)$^2$ Ta-based STJs that were developed for synchrotron measurements (Fig.~\ref{Detectors} (middle)).  This will provide two orders of magnitude increase in statistics over the Phase-II limits for the same running period.  More importantly, these measurements will serve as a test ground to understand any scaling-related limitations before Phase-IV.

\subsection{\label{PhaseIV}Phase-IV: Final Exclusion Limits with High-Resolution Al-Based STJ Arrays}
For Phase-IV, we have started to fabricate and operate arrays of pure Al (200~nm) - Al$_2$O$_3$ (1~nm) - Al (200~nm) STJs.  Al-based STJs have the advantage that the energy gap $\Delta\sim 0.17$~meV is $4\times$ smaller than the energy gap in Ta, potentially providing an intrinsic energy resolution of $\sim 0.5$~eV.  Three arrays will be used with pixel sizes of $(70\mu$m)$^2$, $(130\mu$m)$^2$, and $(200\mu$m)$^2$, respectively, initially allowing us to better understand any in-medium effects as well as compare data across new detectors.  In our current design, the new STJs will be deposited onto $\sim 1~\mu$m thin Si$_3$N$_4$ membranes with the area between pixels covered by a layer of Au to thermalize high-energy substrate phonons to minimize the number of excess charges in the STJs resulting from this background process.  The final limits from the BeEST experiment will result from a 128-pixel Al-STJ array counting at 1 000 counts/s/pixel for 100 days.

\section{Conclusions}
The BeEST experiment is a sensitive, model-independent search for any BSM physics that couples to the neutrino mass using the EC decay of $^7$Be implanted in STJs.  The high-rate ability of STJs allows for the collection of high statistics in reasonable experimental periods, and takes advantage of the high-dose of $^7$Be that is achievable at TRIUMF-ISAC.  The BeEST experiment has completed the first two phases of the project - including setting the first model-independent limits on the existence of HNLs in the 100 - 860 keV mass range from EC decay - and is currently running Phase-III tests.  The demonstration of low-energy nuclear recoil measurements in the first two phases of the BeEST experiment may also initiate new experiments with other rare isotopes in our continued search for BSM physics from nuclear decay.

\begin{acknowledgements}
The BeEST is supported by the U.S. Department of Energy, Office of Science under grant No. DE-SC0021245, and the LLNL Laboratory Directed Research and Development program through grants No. 19-FS-027 and No. 20-LW-006.  TRIUMF receives federal funding via a contribution agreement with the National Research Council of Canada. This work is performed under the auspices of the U.S. Department of Energy by Lawrence Livermore National Laboratory under Contract No. DE-AC52-07NA27344.  We thank the ongoing efforts of the entire BeEST collaboration and acknowledge the extensive efforts that everyone has put into our program over the past several years!
\end{acknowledgements}

\pagebreak

\bibliography{references}

\begin{thebibliography}{25}%
\makeatletter
\providecommand \@ifxundefined [1]{%
 \@ifx{#1\undefined}
}%
\providecommand \@ifnum [1]{%
 \ifnum #1\expandafter \@firstoftwo
 \else \expandafter \@secondoftwo
 \fi
}%
\providecommand \@ifx [1]{%
 \ifx #1\expandafter \@firstoftwo
 \else \expandafter \@secondoftwo
 \fi
}%
\providecommand \natexlab [1]{#1}%
\providecommand \enquote  [1]{``#1''}%
\providecommand \bibnamefont  [1]{#1}%
\providecommand \bibfnamefont [1]{#1}%
\providecommand \citenamefont [1]{#1}%
\providecommand \href@noop [0]{\@secondoftwo}%
\providecommand \href [0]{\begingroup \@sanitize@url \@href}%
\providecommand \@href[1]{\@@startlink{#1}\@@href}%
\providecommand \@@href[1]{\endgroup#1\@@endlink}%
\providecommand \@sanitize@url [0]{\catcode `\\12\catcode `\$12\catcode
  `\&12\catcode `\#12\catcode `\^12\catcode `\_12\catcode `\%12\relax}%
\providecommand \@@startlink[1]{}%
\providecommand \@@endlink[0]{}%
\providecommand \url  [0]{\begingroup\@sanitize@url \@url }%
\providecommand \@url [1]{\endgroup\@href {#1}{\urlprefix }}%
\providecommand \urlprefix  [0]{URL }%
\providecommand \Eprint [0]{\href }%
\providecommand \doibase [0]{http://dx.doi.org/}%
\providecommand \selectlanguage [0]{\@gobble}%
\providecommand \bibinfo  [0]{\@secondoftwo}%
\providecommand \bibfield  [0]{\@secondoftwo}%
\providecommand \translation [1]{[#1]}%
\providecommand \BibitemOpen [0]{}%
\providecommand \bibitemStop [0]{}%
\providecommand \bibitemNoStop [0]{.\EOS\space}%
\providecommand \EOS [0]{\spacefactor3000\relax}%
\providecommand \BibitemShut  [1]{\csname bibitem#1\endcsname}%
\let\auto@bib@innerbib\@empty
%</preamble>
\bibitem [{\citenamefont {Ahmad}\ \emph {et~al.}(2001)\citenamefont {Ahmad}
  \emph {et~al.}}]{Ahm01}%
  \BibitemOpen
  \bibfield  {author} {\bibinfo {author} {\bibfnamefont {Q.~R.}\ \bibnamefont
  {Ahmad}} \emph {et~al.} (\bibinfo {collaboration} {SNO Collaboration}),\
  }\href {\doibase 10.1103/PhysRevLett.87.071301} {\bibfield  {journal}
  {\bibinfo  {journal} {Phys. Rev. Lett.}\ }\textbf {\bibinfo {volume} {87}},\
  \bibinfo {pages} {071301} (\bibinfo {year} {2001})}\BibitemShut {NoStop}%
\bibitem [{\citenamefont {Fukuda}\ \emph {et~al.}(1998)\citenamefont {Fukuda}
  \emph {et~al.}}]{Fuk98}%
  \BibitemOpen
  \bibfield  {author} {\bibinfo {author} {\bibfnamefont {Y.}~\bibnamefont
  {Fukuda}} \emph {et~al.} (\bibinfo {collaboration} {Super-Kamiokande
  Collaboration}),\ }\href {\doibase 10.1103/PhysRevLett.81.1562} {\bibfield
  {journal} {\bibinfo  {journal} {Phys. Rev. Lett.}\ }\textbf {\bibinfo
  {volume} {81}},\ \bibinfo {pages} {1562} (\bibinfo {year}
  {1998})}\BibitemShut {NoStop}%
\bibitem [{\citenamefont {Abi}\ \emph {et~al.}(2021)\citenamefont {Abi} \emph
  {et~al.}}]{Abi21}%
  \BibitemOpen
  \bibfield  {author} {\bibinfo {author} {\bibfnamefont {B.}~\bibnamefont
  {Abi}} \emph {et~al.} (\bibinfo {collaboration} {Muon $g\ensuremath{-}2$
  Collaboration}),\ }\href {\doibase 10.1103/PhysRevLett.126.141801} {\bibfield
   {journal} {\bibinfo  {journal} {Phys. Rev. Lett.}\ }\textbf {\bibinfo
  {volume} {126}},\ \bibinfo {pages} {141801} (\bibinfo {year}
  {2021})}\BibitemShut {NoStop}%
\bibitem [{\citenamefont {Garisto}(2020)}]{Gar20}%
  \BibitemOpen
  \bibfield  {author} {\bibinfo {author} {\bibfnamefont {D.}~\bibnamefont
  {Garisto}},\ }\href@noop {} {\bibfield  {journal} {\bibinfo  {journal}
  {Physics}\ }\textbf {\bibinfo {volume} {13}},\ \bibinfo {pages} {79}
  (\bibinfo {year} {2020})}\BibitemShut {NoStop}%
\bibitem [{\citenamefont {Boyarsky}\ \emph {et~al.}(2019)\citenamefont
  {Boyarsky}, \citenamefont {Drewes}, \citenamefont {Lasserre}, \citenamefont
  {Mertens},\ and\ \citenamefont {Ruchayskiy}}]{Boy19}%
  \BibitemOpen
  \bibfield  {author} {\bibinfo {author} {\bibfnamefont {A.}~\bibnamefont
  {Boyarsky}}, \bibinfo {author} {\bibfnamefont {M.}~\bibnamefont {Drewes}},
  \bibinfo {author} {\bibfnamefont {T.}~\bibnamefont {Lasserre}}, \bibinfo
  {author} {\bibfnamefont {S.}~\bibnamefont {Mertens}}, \ and\ \bibinfo
  {author} {\bibfnamefont {O.}~\bibnamefont {Ruchayskiy}},\ }\href {\doibase
  https://doi.org/10.1016/j.ppnp.2018.07.004} {\bibfield  {journal} {\bibinfo
  {journal} {Progress in Particle and Nuclear Physics}\ }\textbf {\bibinfo
  {volume} {104}},\ \bibinfo {pages} {1 } (\bibinfo {year} {2019})}\BibitemShut
  {NoStop}%
\bibitem [{\citenamefont {Adhikari}\ \emph {et~al.}(2017)\citenamefont
  {Adhikari} \emph {et~al.}}]{Adh17}%
  \BibitemOpen
  \bibfield  {author} {\bibinfo {author} {\bibfnamefont {R.}~\bibnamefont
  {Adhikari}} \emph {et~al.},\ }\href {\doibase 10.1088/1475-7516/2017/01/025}
  {\bibfield  {journal} {\bibinfo  {journal} {Journal of Cosmology and
  Astroparticle Physics}\ }\textbf {\bibinfo {volume} {2017}},\ \bibinfo
  {pages} {025} (\bibinfo {year} {2017})}\BibitemShut {NoStop}%
\bibitem [{\citenamefont {Asaka}\ \emph {et~al.}(2005)\citenamefont {Asaka},
  \citenamefont {Blanchet},\ and\ \citenamefont {Shaposhnikov}}]{Asa05a}%
  \BibitemOpen
  \bibfield  {author} {\bibinfo {author} {\bibfnamefont {T.}~\bibnamefont
  {Asaka}}, \bibinfo {author} {\bibfnamefont {S.}~\bibnamefont {Blanchet}}, \
  and\ \bibinfo {author} {\bibfnamefont {M.}~\bibnamefont {Shaposhnikov}},\
  }\href {\doibase https://doi.org/10.1016/j.physletb.2005.09.070} {\bibfield
  {journal} {\bibinfo  {journal} {Physics Letters B}\ }\textbf {\bibinfo
  {volume} {631}},\ \bibinfo {pages} {151 } (\bibinfo {year}
  {2005})}\BibitemShut {NoStop}%
\bibitem [{\citenamefont {Asaka}\ and\ \citenamefont
  {Shaposhnikov}(2005)}]{Asa05b}%
  \BibitemOpen
  \bibfield  {author} {\bibinfo {author} {\bibfnamefont {T.}~\bibnamefont
  {Asaka}}\ and\ \bibinfo {author} {\bibfnamefont {M.}~\bibnamefont
  {Shaposhnikov}},\ }\href {\doibase
  https://doi.org/10.1016/j.physletb.2005.06.020} {\bibfield  {journal}
  {\bibinfo  {journal} {Physics Letters B}\ }\textbf {\bibinfo {volume}
  {620}},\ \bibinfo {pages} {17 } (\bibinfo {year} {2005})}\BibitemShut
  {NoStop}%
\bibitem [{\citenamefont {Wang}\ \emph {et~al.}(2017)\citenamefont {Wang},
  \citenamefont {Audi}, \citenamefont {Kondev}, \citenamefont {Huang},
  \citenamefont {Naimi},\ and\ \citenamefont {Xu}}]{AME16}%
  \BibitemOpen
  \bibfield  {author} {\bibinfo {author} {\bibfnamefont {M.}~\bibnamefont
  {Wang}}, \bibinfo {author} {\bibfnamefont {G.}~\bibnamefont {Audi}}, \bibinfo
  {author} {\bibfnamefont {F.}~\bibnamefont {Kondev}}, \bibinfo {author}
  {\bibfnamefont {W.}~\bibnamefont {Huang}}, \bibinfo {author} {\bibfnamefont
  {S.}~\bibnamefont {Naimi}}, \ and\ \bibinfo {author} {\bibfnamefont
  {X.}~\bibnamefont {Xu}},\ }\href
  {http://stacks.iop.org/1674-1137/41/i=3/a=030003} {\bibfield  {journal}
  {\bibinfo  {journal} {Chinese Physics C}\ }\textbf {\bibinfo {volume} {41}},\
  \bibinfo {pages} {030003} (\bibinfo {year} {2017})}\BibitemShut {NoStop}%
\bibitem [{\citenamefont {Tilley}\ \emph {et~al.}(2002)\citenamefont {Tilley},
  \citenamefont {Cheves}, \citenamefont {Godwin}, \citenamefont {Hale},
  \citenamefont {Hofmann}, \citenamefont {Kelley}, \citenamefont {Sheu},\ and\
  \citenamefont {Weller}}]{Til02}%
  \BibitemOpen
  \bibfield  {author} {\bibinfo {author} {\bibfnamefont {D.}~\bibnamefont
  {Tilley}}, \bibinfo {author} {\bibfnamefont {C.}~\bibnamefont {Cheves}},
  \bibinfo {author} {\bibfnamefont {J.}~\bibnamefont {Godwin}}, \bibinfo
  {author} {\bibfnamefont {G.}~\bibnamefont {Hale}}, \bibinfo {author}
  {\bibfnamefont {H.}~\bibnamefont {Hofmann}}, \bibinfo {author} {\bibfnamefont
  {J.}~\bibnamefont {Kelley}}, \bibinfo {author} {\bibfnamefont
  {C.}~\bibnamefont {Sheu}}, \ and\ \bibinfo {author} {\bibfnamefont
  {H.}~\bibnamefont {Weller}},\ }\href {\doibase
  https://doi.org/10.1016/S0375-9474(02)00597-3} {\bibfield  {journal}
  {\bibinfo  {journal} {Nuclear Physics A}\ }\textbf {\bibinfo {volume}
  {708}},\ \bibinfo {pages} {3 } (\bibinfo {year} {2002})}\BibitemShut
  {NoStop}%
\bibitem [{\citenamefont {Helmer}\ and\ \citenamefont {van~der
  Leun}(2000)}]{Hel00}%
  \BibitemOpen
  \bibfield  {author} {\bibinfo {author} {\bibfnamefont {R.}~\bibnamefont
  {Helmer}}\ and\ \bibinfo {author} {\bibfnamefont {C.}~\bibnamefont {van~der
  Leun}},\ }\href {\doibase https://doi.org/10.1016/S0168-9002(00)00252-7}
  {\bibfield  {journal} {\bibinfo  {journal} {Nuclear Instruments and Methods
  in Physics Research Section A: Accelerators, Spectrometers, Detectors and
  Associated Equipment}\ }\textbf {\bibinfo {volume} {450}},\ \bibinfo {pages}
  {35 } (\bibinfo {year} {2000})}\BibitemShut {NoStop}%
\bibitem [{\citenamefont {Voytas}\ \emph {et~al.}(2001)\citenamefont {Voytas},
  \citenamefont {Ternovan}, \citenamefont {Galeazzi}, \citenamefont {McCammon},
  \citenamefont {Kolata}, \citenamefont {Santi}, \citenamefont {Peterson},
  \citenamefont {Guimar\~aes}, \citenamefont {Becchetti}, \citenamefont {Lee},
  \citenamefont {O'Donnell}, \citenamefont {Roberts},\ and\ \citenamefont
  {Shaheen}}]{Voy02}%
  \BibitemOpen
  \bibfield  {author} {\bibinfo {author} {\bibfnamefont {P.~A.}\ \bibnamefont
  {Voytas}}, \bibinfo {author} {\bibfnamefont {C.}~\bibnamefont {Ternovan}},
  \bibinfo {author} {\bibfnamefont {M.}~\bibnamefont {Galeazzi}}, \bibinfo
  {author} {\bibfnamefont {D.}~\bibnamefont {McCammon}}, \bibinfo {author}
  {\bibfnamefont {J.~J.}\ \bibnamefont {Kolata}}, \bibinfo {author}
  {\bibfnamefont {P.}~\bibnamefont {Santi}}, \bibinfo {author} {\bibfnamefont
  {D.}~\bibnamefont {Peterson}}, \bibinfo {author} {\bibfnamefont
  {V.}~\bibnamefont {Guimar\~aes}}, \bibinfo {author} {\bibfnamefont {F.~D.}\
  \bibnamefont {Becchetti}}, \bibinfo {author} {\bibfnamefont {M.~Y.}\
  \bibnamefont {Lee}}, \bibinfo {author} {\bibfnamefont {T.~W.}\ \bibnamefont
  {O'Donnell}}, \bibinfo {author} {\bibfnamefont {D.~A.}\ \bibnamefont
  {Roberts}}, \ and\ \bibinfo {author} {\bibfnamefont {S.}~\bibnamefont
  {Shaheen}},\ }\href {\doibase 10.1103/PhysRevLett.88.012501} {\bibfield
  {journal} {\bibinfo  {journal} {Phys. Rev. Lett.}\ }\textbf {\bibinfo
  {volume} {88}},\ \bibinfo {pages} {012501} (\bibinfo {year}
  {2001})}\BibitemShut {NoStop}%
\bibitem [{\citenamefont {Fretwell}\ \emph {et~al.}(2020)\citenamefont
  {Fretwell}, \citenamefont {Leach}, \citenamefont {Bray}, \citenamefont {Kim},
  \citenamefont {Dilling}, \citenamefont {Lennarz}, \citenamefont {Mougeot},
  \citenamefont {Ponce}, \citenamefont {Ruiz}, \citenamefont {Stackhouse},\
  and\ \citenamefont {Friedrich}}]{Fre20}%
  \BibitemOpen
  \bibfield  {author} {\bibinfo {author} {\bibfnamefont {S.}~\bibnamefont
  {Fretwell}}, \bibinfo {author} {\bibfnamefont {K.~G.}\ \bibnamefont {Leach}},
  \bibinfo {author} {\bibfnamefont {C.}~\bibnamefont {Bray}}, \bibinfo {author}
  {\bibfnamefont {G.~B.}\ \bibnamefont {Kim}}, \bibinfo {author} {\bibfnamefont
  {J.}~\bibnamefont {Dilling}}, \bibinfo {author} {\bibfnamefont
  {A.}~\bibnamefont {Lennarz}}, \bibinfo {author} {\bibfnamefont
  {X.}~\bibnamefont {Mougeot}}, \bibinfo {author} {\bibfnamefont
  {F.}~\bibnamefont {Ponce}}, \bibinfo {author} {\bibfnamefont
  {C.}~\bibnamefont {Ruiz}}, \bibinfo {author} {\bibfnamefont {J.}~\bibnamefont
  {Stackhouse}}, \ and\ \bibinfo {author} {\bibfnamefont {S.}~\bibnamefont
  {Friedrich}},\ }\href {\doibase 10.1103/PhysRevLett.125.032701} {\bibfield
  {journal} {\bibinfo  {journal} {Phys. Rev. Lett.}\ }\textbf {\bibinfo
  {volume} {125}},\ \bibinfo {pages} {032701} (\bibinfo {year}
  {2020})}\BibitemShut {NoStop}%
\bibitem [{\citenamefont {Kurakado}(1982)}]{Kur82}%
  \BibitemOpen
  \bibfield  {author} {\bibinfo {author} {\bibfnamefont {M.}~\bibnamefont
  {Kurakado}},\ }\href {\doibase https://doi.org/10.1016/0029-554X(82)90654-1}
  {\bibfield  {journal} {\bibinfo  {journal} {Nuclear Instruments and Methods
  in Physics Research}\ }\textbf {\bibinfo {volume} {196}},\ \bibinfo {pages}
  {275 } (\bibinfo {year} {1982})}\BibitemShut {NoStop}%
\bibitem [{\citenamefont {{Friedrich}}\ \emph {et~al.}(2003)\citenamefont
  {{Friedrich}}, \citenamefont {{Vailionis}}, \citenamefont {{Drury}},
  \citenamefont {{Niedermayr}}, \citenamefont {{Funk}}, \citenamefont {{Kang}},
  \citenamefont {{Eun-Mi Choi}}, \citenamefont {{Hyeong-Jin Kim}},
  \citenamefont {{Sung-Ik Lee}}, \citenamefont {{Cramer}}, \citenamefont
  {{Changyoung Kim}},\ and\ \citenamefont {{Labov}}}]{Fri03}%
  \BibitemOpen
  \bibfield  {author} {\bibinfo {author} {\bibfnamefont {S.}~\bibnamefont
  {{Friedrich}}}, \bibinfo {author} {\bibfnamefont {A.}~\bibnamefont
  {{Vailionis}}}, \bibinfo {author} {\bibfnamefont {O.}~\bibnamefont
  {{Drury}}}, \bibinfo {author} {\bibfnamefont {T.}~\bibnamefont
  {{Niedermayr}}}, \bibinfo {author} {\bibfnamefont {T.}~\bibnamefont
  {{Funk}}}, \bibinfo {author} {\bibfnamefont {W.~N.}\ \bibnamefont {{Kang}}},
  \bibinfo {author} {\bibnamefont {{Eun-Mi Choi}}}, \bibinfo {author}
  {\bibnamefont {{Hyeong-Jin Kim}}}, \bibinfo {author} {\bibnamefont {{Sung-Ik
  Lee}}}, \bibinfo {author} {\bibfnamefont {S.~P.}\ \bibnamefont {{Cramer}}},
  \bibinfo {author} {\bibnamefont {{Changyoung Kim}}}, \ and\ \bibinfo {author}
  {\bibfnamefont {S.~E.}\ \bibnamefont {{Labov}}},\ }\href {\doibase
  10.1109/TASC.2003.814169} {\bibfield  {journal} {\bibinfo  {journal} {IEEE
  Transactions on Applied Superconductivity}\ }\textbf {\bibinfo {volume}
  {13}},\ \bibinfo {pages} {1114} (\bibinfo {year} {2003})}\BibitemShut
  {NoStop}%
\bibitem [{\citenamefont {Verhoeve}(2008)}]{Ver08}%
  \BibitemOpen
  \bibfield  {author} {\bibinfo {author} {\bibfnamefont {P.}~\bibnamefont
  {Verhoeve}},\ }\href {\doibase 10.1007/s10909-008-9730-9} {\bibfield
  {journal} {\bibinfo  {journal} {Journal of Low Temperature Physics}\ }\textbf
  {\bibinfo {volume} {151}},\ \bibinfo {pages} {675} (\bibinfo {year}
  {2008})}\BibitemShut {NoStop}%
\bibitem [{\citenamefont {Carpenter}\ \emph {et~al.}(2014)\citenamefont
  {Carpenter}, \citenamefont {Friedrich}, \citenamefont {Hall}, \citenamefont
  {Harris},\ and\ \citenamefont {Cantor}}]{Car14}%
  \BibitemOpen
  \bibfield  {author} {\bibinfo {author} {\bibfnamefont {M.~H.}\ \bibnamefont
  {Carpenter}}, \bibinfo {author} {\bibfnamefont {S.}~\bibnamefont
  {Friedrich}}, \bibinfo {author} {\bibfnamefont {J.~A.}\ \bibnamefont {Hall}},
  \bibinfo {author} {\bibfnamefont {J.}~\bibnamefont {Harris}}, \ and\ \bibinfo
  {author} {\bibfnamefont {R.}~\bibnamefont {Cantor}},\ }\href {\doibase
  10.1007/s10909-014-1172-y} {\bibfield  {journal} {\bibinfo  {journal}
  {Journal of Low Temperature Physics}\ }\textbf {\bibinfo {volume} {176}},\
  \bibinfo {pages} {222} (\bibinfo {year} {2014})}\BibitemShut {NoStop}%
\bibitem [{\citenamefont {Ponce}\ \emph {et~al.}(2016)\citenamefont {Ponce},
  \citenamefont {Carpenter}, \citenamefont {Cantor},\ and\ \citenamefont
  {Friedrich}}]{Pon16}%
  \BibitemOpen
  \bibfield  {author} {\bibinfo {author} {\bibfnamefont {F.}~\bibnamefont
  {Ponce}}, \bibinfo {author} {\bibfnamefont {M.~H.}\ \bibnamefont
  {Carpenter}}, \bibinfo {author} {\bibfnamefont {R.}~\bibnamefont {Cantor}}, \
  and\ \bibinfo {author} {\bibfnamefont {S.}~\bibnamefont {Friedrich}},\ }\href
  {\doibase 10.1007/s10909-015-1443-2} {\bibfield  {journal} {\bibinfo
  {journal} {Journal of Low Temperature Physics}\ }\textbf {\bibinfo {volume}
  {184}},\ \bibinfo {pages} {694} (\bibinfo {year} {2016})}\BibitemShut
  {NoStop}%
\bibitem [{\citenamefont {Friedrich}\ \emph {et~al.}(2020)\citenamefont
  {Friedrich}, \citenamefont {Ponce}, \citenamefont {Hall},\ and\ \citenamefont
  {Cantor}}]{Fri20}%
  \BibitemOpen
  \bibfield  {author} {\bibinfo {author} {\bibfnamefont {S.}~\bibnamefont
  {Friedrich}}, \bibinfo {author} {\bibfnamefont {F.}~\bibnamefont {Ponce}},
  \bibinfo {author} {\bibfnamefont {J.~A.}\ \bibnamefont {Hall}}, \ and\
  \bibinfo {author} {\bibfnamefont {R.}~\bibnamefont {Cantor}},\ }\href
  {\doibase 10.1007/s10909-020-02360-2} {\bibfield  {journal} {\bibinfo
  {journal} {Journal of Low Temperature Physics}\ }\textbf {\bibinfo {volume}
  {200}},\ \bibinfo {pages} {200} (\bibinfo {year} {2020})}\BibitemShut
  {NoStop}%
\bibitem [{\citenamefont {{Friedrich}}\ \emph {et~al.}(2009)\citenamefont
  {{Friedrich}}, \citenamefont {{Hertrich}}, \citenamefont {{Drury}},
  \citenamefont {{Cherepy}},\ and\ \citenamefont {{Hohne}}}]{Fri09}%
  \BibitemOpen
  \bibfield  {author} {\bibinfo {author} {\bibfnamefont {S.}~\bibnamefont
  {{Friedrich}}}, \bibinfo {author} {\bibfnamefont {T.}~\bibnamefont
  {{Hertrich}}}, \bibinfo {author} {\bibfnamefont {O.~B.}\ \bibnamefont
  {{Drury}}}, \bibinfo {author} {\bibfnamefont {N.~J.}\ \bibnamefont
  {{Cherepy}}}, \ and\ \bibinfo {author} {\bibfnamefont {J.}~\bibnamefont
  {{Hohne}}},\ }\href {\doibase 10.1109/TNS.2009.2014062} {\bibfield  {journal}
  {\bibinfo  {journal} {IEEE Transactions on Nuclear Science}\ }\textbf
  {\bibinfo {volume} {56}},\ \bibinfo {pages} {1089} (\bibinfo {year}
  {2009})}\BibitemShut {NoStop}%
\bibitem [{\citenamefont {Friedrich}\ \emph {et~al.}(2021)\citenamefont
  {Friedrich}, \citenamefont {Kim}, \citenamefont {Bray}, \citenamefont
  {Cantor}, \citenamefont {Dilling}, \citenamefont {Fretwell}, \citenamefont
  {Hall}, \citenamefont {Lennarz}, \citenamefont {Lordi}, \citenamefont
  {Machule}, \citenamefont {McKeen}, \citenamefont {Mougeot}, \citenamefont
  {Ponce}, \citenamefont {Ruiz}, \citenamefont {Samanta}, \citenamefont
  {Warburton},\ and\ \citenamefont {Leach}}]{Fri21}%
  \BibitemOpen
  \bibfield  {author} {\bibinfo {author} {\bibfnamefont {S.}~\bibnamefont
  {Friedrich}}, \bibinfo {author} {\bibfnamefont {G.~B.}\ \bibnamefont {Kim}},
  \bibinfo {author} {\bibfnamefont {C.}~\bibnamefont {Bray}}, \bibinfo {author}
  {\bibfnamefont {R.}~\bibnamefont {Cantor}}, \bibinfo {author} {\bibfnamefont
  {J.}~\bibnamefont {Dilling}}, \bibinfo {author} {\bibfnamefont
  {S.}~\bibnamefont {Fretwell}}, \bibinfo {author} {\bibfnamefont {J.~A.}\
  \bibnamefont {Hall}}, \bibinfo {author} {\bibfnamefont {A.}~\bibnamefont
  {Lennarz}}, \bibinfo {author} {\bibfnamefont {V.}~\bibnamefont {Lordi}},
  \bibinfo {author} {\bibfnamefont {P.}~\bibnamefont {Machule}}, \bibinfo
  {author} {\bibfnamefont {D.}~\bibnamefont {McKeen}}, \bibinfo {author}
  {\bibfnamefont {X.}~\bibnamefont {Mougeot}}, \bibinfo {author} {\bibfnamefont
  {F.}~\bibnamefont {Ponce}}, \bibinfo {author} {\bibfnamefont
  {C.}~\bibnamefont {Ruiz}}, \bibinfo {author} {\bibfnamefont {A.}~\bibnamefont
  {Samanta}}, \bibinfo {author} {\bibfnamefont {W.~K.}\ \bibnamefont
  {Warburton}}, \ and\ \bibinfo {author} {\bibfnamefont {K.~G.}\ \bibnamefont
  {Leach}},\ }\href {\doibase 10.1103/PhysRevLett.126.021803} {\bibfield
  {journal} {\bibinfo  {journal} {Phys. Rev. Lett.}\ }\textbf {\bibinfo
  {volume} {126}},\ \bibinfo {pages} {021803} (\bibinfo {year}
  {2021})}\BibitemShut {NoStop}%
\bibitem [{\citenamefont {Dilling}\ and\ \citenamefont
  {Kr{\"u}cken}(2014)}]{Dil14}%
  \BibitemOpen
  \bibfield  {author} {\bibinfo {author} {\bibfnamefont {J.}~\bibnamefont
  {Dilling}}\ and\ \bibinfo {author} {\bibfnamefont {R.}~\bibnamefont
  {Kr{\"u}cken}},\ }\href {\doibase 10.1007/s10751-013-0886-6} {\bibfield
  {journal} {\bibinfo  {journal} {Hyperfine Interactions}\ }\textbf {\bibinfo
  {volume} {225}},\ \bibinfo {pages} {111} (\bibinfo {year}
  {2014})}\BibitemShut {NoStop}%
\bibitem [{\citenamefont {Blumenfeld}\ \emph {et~al.}(2013)\citenamefont
  {Blumenfeld}, \citenamefont {Nilsson},\ and\ \citenamefont {Duppen}}]{Blu13}%
  \BibitemOpen
  \bibfield  {author} {\bibinfo {author} {\bibfnamefont {Y.}~\bibnamefont
  {Blumenfeld}}, \bibinfo {author} {\bibfnamefont {T.}~\bibnamefont {Nilsson}},
  \ and\ \bibinfo {author} {\bibfnamefont {P.~V.}\ \bibnamefont {Duppen}},\
  }\href {http://stacks.iop.org/1402-4896/2013/i=T152/a=014023} {\bibfield
  {journal} {\bibinfo  {journal} {Physica Scripta}\ }\textbf {\bibinfo {volume}
  {2013}},\ \bibinfo {pages} {014023} (\bibinfo {year} {2013})}\BibitemShut
  {NoStop}%
\bibitem [{\citenamefont {{Carpenter}}\ \emph {et~al.}(2013)\citenamefont
  {{Carpenter}}, \citenamefont {{Friedrich}}, \citenamefont {{Hall}},
  \citenamefont {{Harris}}, \citenamefont {{Warburton}},\ and\ \citenamefont
  {{Cantor}}}]{Car13}%
  \BibitemOpen
  \bibfield  {author} {\bibinfo {author} {\bibfnamefont {M.~H.}\ \bibnamefont
  {{Carpenter}}}, \bibinfo {author} {\bibfnamefont {S.}~\bibnamefont
  {{Friedrich}}}, \bibinfo {author} {\bibfnamefont {J.~A.}\ \bibnamefont
  {{Hall}}}, \bibinfo {author} {\bibfnamefont {J.}~\bibnamefont {{Harris}}},
  \bibinfo {author} {\bibfnamefont {W.~K.}\ \bibnamefont {{Warburton}}}, \ and\
  \bibinfo {author} {\bibfnamefont {R.}~\bibnamefont {{Cantor}}},\ }\href
  {\doibase 10.1109/TASC.2012.2236877} {\bibfield  {journal} {\bibinfo
  {journal} {IEEE Transactions on Applied Superconductivity}\ }\textbf
  {\bibinfo {volume} {23}},\ \bibinfo {pages} {2400504} (\bibinfo {year}
  {2013})}\BibitemShut {NoStop}%
\bibitem [{\citenamefont {Ponce}\ \emph {et~al.}(2018)\citenamefont {Ponce},
  \citenamefont {Swanberg}, \citenamefont {Burke}, \citenamefont {Henderson},\
  and\ \citenamefont {Friedrich}}]{Pon18}%
  \BibitemOpen
  \bibfield  {author} {\bibinfo {author} {\bibfnamefont {F.}~\bibnamefont
  {Ponce}}, \bibinfo {author} {\bibfnamefont {E.}~\bibnamefont {Swanberg}},
  \bibinfo {author} {\bibfnamefont {J.}~\bibnamefont {Burke}}, \bibinfo
  {author} {\bibfnamefont {R.}~\bibnamefont {Henderson}}, \ and\ \bibinfo
  {author} {\bibfnamefont {S.}~\bibnamefont {Friedrich}},\ }\href {\doibase
  10.1103/PhysRevC.97.054310} {\bibfield  {journal} {\bibinfo  {journal} {Phys.
  Rev. C}\ }\textbf {\bibinfo {volume} {97}},\ \bibinfo {pages} {054310}
  (\bibinfo {year} {2018})}\BibitemShut {NoStop}%
\end{thebibliography}%

\end{document}